\documentstyle[aps,preprint,prl]{revtex}
\begin{document}
\draft
\title{Theory of adsorption and surfactant effect of Sb on Ag (111)}
\author{Sabrina Oppo,\cite{p_add}
	Vincenzo Fiorentini,\cite{p_add} and
	Matthias Scheffler}
\address{Fritz-Haber-Institut der Max-Planck-Gesellschaft,
 	  Faradayweg 4-6, D-14195 Berlin, Germany}
\date{Submitted to Physical Review Letters on 13 July 1993}
\maketitle
\begin{abstract}
We present first-principles studies of the adsorption of Sb and Ag on clean and
Sb-covered  Ag (111).
 For Sb, the {\it substitutional} adsorption  site is found to be
greatly favored with respect to  on-surface fcc sites and to  subsurface
 sites, so that a  segregating  surface alloy layer is formed. Adsorbed
silver adatoms are  more strongly bound on clean Ag(111) than on Sb-covered Ag.
 We propose that the experimentally reported surfactant effect of Sb is due
to  Sb adsorbates reducing the Ag adatom mobility. This gives rise to a
high density of Ag islands which  coalesce into regular layers.
\end{abstract}
\narrowtext

The  goal of epitaxial crystal growth is to achieve
 atomically-flat and defect-free surfaces of specified crystallographic
orientation, under the widest possible range of growth conditions.
Significant  efforts are devoted since many years to the growth of
semiconductors. The epitaxial growth of metals on metal substrates has also
attracted considerable  interest (see for example Ref.\cite{mgrowth} and
references therein).

Layer-by-layer, or two-dimensional (2D), growth is such that
 the epitaxial layer being currently deposited is completed before
further layers begin to grow on top of it; this mode is also named
Frank-van der Merwe. In the three-dimensional (3D) or cluster growth,
many overlayers grow at the same time, none of them being completed, so that
the surface exhibits 3D islands. For heteroepitaxy, depending on
whether the 3D mode manifests itself immediately or only after the
formation of a few 2D overlayers, the  3D growth is named either
after  Volmer and Weber, or Stransky and Krastanov.

With a view  at extending the external conditions for the growth of
high-quality 2D surfaces towards lower temperatures and higher deposition
rates, use has been made recently of surface contaminants which purportedly
act as  surfactants. Although by definition\cite{oxf}  a surfactant should
reduce the surface  formation energy\cite{ssurfact},
 there is at present no consensus as to  the actual mechanisms of surfactant
action, {\it e.g.} as to  whether the contaminant affects the surface
 energy or the kinetics of growth\cite{mass}; the term surfactant
is thus often used in the broader sense that it promotes
 2D growth as opposed to 3D growth.

The surfactant technique, although still
in its infancy, has been by now rather widely applied\cite{ssurfact,mass}
in the field of semiconductors to help regular growth of heteroepitaxial
strained layers. On the other hand, we are only aware of one report
of surfactant-assisted growth of metals\cite{vdv}. This is concerned with
the homoepitaxy of the (111) surface of Ag. The growth mode was found  to
be drastically altered, from  3D to layer-by-layer 2D, by the one-time
deposition of Sb at the beginning
of the growth process, at coverages $\Theta$ between
0.05 and 0.2. Clean  Ag (111) was
 observed to  grow in a 3D fashion between  250 and 400 K,
as signaled by the exponential decrease of the reflected x-ray beam intensity
which monitors the degree of coherence of the upper layers of the sample
\cite{vdv}. A crossover to step-flow growth (corresponding to
constant reflected intensity)  was observed above  450-500 K.
In the presence of Sb, an  oscillatory behaviour of the
reflected intensity was observed instead, which is a  fingerprint of
 2D growth. The layer-by-layer growth of Sb-precovered Ag (111) continues
for a  rather long time
(typically equivalent to the growth of  25 monolayers or more),
 at a nominal Ag deposition rate of 0.02 monolayer per second,
even at 280-300 K.

The actual growth mode of an ideal  clean Ag surface is still unclear. Indeed,
Ag would be expected to grow layer-by-layer even at rather low  temperatures;
3D cluster  growth might be initiated {\it e.g.} by nucleation at surface
defects. This notwithstanding, Sb unambiguously promotes the 2D growth of Ag,
and it is therefore important to  investigate the pertaining mechanism. With
this aim, we have performed   {\it ab initio} studies of the energetics
of Sb and Ag adsorption on Ag (111). The calculations presented in this Letter
are performed at  Sb coverages down to  $\Theta=1/4$.
Among the  considered adsorption geometries,
 the most stable one is the substitutional site, Sb being bound
into a Ag surface vacancy. This  site is considerably  more favourable than the
conventional on-surface fcc sites, and than subsurface
sites. Thus, an Sb-Ag alloy layer forms in  the surface layer.
Dissolution of adsorbed  Sb   into bulk Ag is
energetically disfavoured. When covered with Ag,  the substitutional
surface alloy  reforms as the topmost  surface layer
by segregation of Sb.
Due to the need of forming surface vacancies, the formation of the
substitutional surface alloy needs thermal activation; we predict
that at the relevant temperature,
 at low enough coverages,  disordering of the surface alloy should
take  place.

Calculations for Ag adsorbed on clean and substitutional Sb-covered Ag (111)
give informations on the growth mode of Ag. The main result is that
Ag is more bound  on clean portions of the  surface, while the vicinity
of substitutional Sb centers is less favorable.
This implies that the average barrier for Ag diffusion is increased, thus
Sb reduces the surface  mobility of Ag. This  causes
a high Ag island density, which reduces the probability of 3D growth.
The segregation of Sb to the surface layer allows the process to
continue.

The calculations were performed within density-functional
theory\cite{gd} in the local density approximation (LDA), using
the all-electron full-potential linear muffin-tin orbitals (LMTO)
method\cite{fp}.
We used a  non-relativistic code, which gives a very  good description
of Ag (bulk
equilibrium properties:  $a_0^{\rm th}=7.73$ bohr, $B^{\rm th}=1.10$
 Mbar, zero-point energy not included,
to be compared with the low-temperature experimental values
 $a_0^{\rm exp}=7.74$
bohr, $B^{\rm exp}=1.01$ Mbar) and Sb, which is  only slightly heavier.
The (111) surface of fcc Ag, clean or with Sb  coverages of   $\Theta$=1,
 $\Theta$=1/3, and $\Theta$=1/4
 (whereby the 1$\times$1, ($\sqrt{3}$$\times$$\sqrt{3}$) $R$30$^{\circ}$, and
2$\times$2 cells
were used respectively), was simulated by slabs of thickness ranging from 5
to 13 atomic layers, separated by 10 layers of vacuum. The  supercells
contained a number of atoms ranging from 7 to 30.
The {\bf k}-summation was done on a uniform mesh
 in the irreducible part of the surface Brillouin zone, encompassing
  19 points for   the 1$\times$1 cell, 7 and  13 points for the
clean and adsorbate-covered $\sqrt{3}\times\sqrt{3}$ cell,
 5 and 9 points for the clean and  adsorbate-covered 2$\times$2 cell.
The vertical position of the adsorbates was optimized.
Substrate  relaxation is neglected, but  is expected to change
 the adsorption energies only marginally. Full details of this study will be
presented elsewhere\cite{noi}.

The binding energy for Sb on-surface fcc adsorption is
$$ E_{\rm ad}^{ \rm fcc} = -(\frac{1}{2}E^{\rm Sb/Ag(111)} -
\frac{1}{2}E^{\rm Ag(111)} -  E^{\rm Sb}_{\rm atom})$$
with $E^{\rm Sb/Ag(111)}$, $E^{\rm Ag(111)}$, and  $E^{\rm Sb}_{\rm atom}$
being  the total
energies of the  adsorbate-covered slab, of the clean Ag slab, and of
the spin-polarized Sb free atom; the factor 1/2 accounts
for the facts that we adsorb on  both slab sides. In
the case of substitutional Sb
adsorption, a slightly different  definition applies:
$$E_{\rm ad}^{ \rm sub} = -[(\frac{1}{2}E^{\rm Sb/Ag(111) sub}
+ E^{\rm Ag}_{\rm bulk}) \\
- (\frac{1}{2}E^{\rm Ag(111)} - E^{\rm Sb}_{\rm atom})]$$
 where $E^{\rm Ag}_{\rm bulk}$ is the bulk total
 energy per atom of fcc Ag. The substitutional process implies in fact that a
surface vacancy
 be created, and the kicked-out Ag atom migrates to a kink site at a
surface step, thus gaining the cohesive energy\cite{neu}. While the formation
of a surface vacancy costs energy, the subsequent binding of the adsorbate
into the vacancy leads to a net energy gain. While at $\Theta$=1 we only have
on-surface adsorption,  substitutional adsorption
with Sb adatoms being not nearest neighbours is possible for
all $\Theta\leq$1/3 (we did not consider the coverage  0.3$<\Theta<$1).

The calculated adsorption energies for the substitutional, fcc on-surface, and
sublayer adsorption are given in Table \ref{tab1} for all
coverages studied. As seen from the Table, the
substitutional site is greatly favoured with respect to ``normal'' on-surface
fcc adsorption, and also against sublayer adsorption. Sb is thus expected
to be adsorbed in substitutional sites\cite{surfe}.

An obstacle to the establishment of a substitutional
adsorbate superstructure is the energy barrier which  may exist for
vacancy formation. To estimate the
barrier, we calculated the formation energy of a
distant Frenkel pair\cite{neu}, consisting of an isolated
 Ag adatom on Ag (111) plus a  vacancy. The results are summarized in  Table
\ref{tab2}. The resulting maximum barrier of about 1.5 eV corresponds
to an activation  temperature of about 500 K. At that temperature
the surface mobility of Ag atoms on Ag (111) is very high, so that migration
of the atom released from the vacancy to a kink site is  easily
achieved.  We note that, if dissipated locally, the adsorption energy
of  Sb in the fcc site would be more than enough to create a
surface vacancy.

As a check as to whether Sb might be incorporated
into the bulk of Ag, we calculated the adsorption energy for
 Sb in a sublayer site, {\it i.e.} below  one overlayer
of Ag. As seen from Table \ref{tab1}, this  site is
strongly  disfavored with respect to Sb sitting in a substitutional
site {\it in} the surface layer. For  Sb  below two Ag overlayers,
the adsorption energy  decreases further.
We  conclude that  Sb  segregates
 to  the surface and is not incorporated into Ag.
Indeed,  the segregation of Sb
(and the ensuing reduction of surface energy) in transition and noble
 metals and alloys has been known for some time in metallurgy\cite{seg}.
In the present case, segregation
 is essentially  due to the size difference of Ag and Sb, as
has been checked by additional calculations at a 5\% increased lattice
constant. For  $\Theta$=1/3, this gives that
 the difference of substitutional  and sublayer
adsorption energies drops from 1.1 eV/atom to about
0.65 eV/atom. Sb is  just about the  right size to fit into a surface vacancy,
but it is somewhat too
large for a bulk vacancy. Since  Sb is   confined   into the
surface layer,  an Sb-Ag alloy layer  will form at
the (111) surface of Ag upon submonolayer Sb deposition. It is worth noticing
that  the substitutional configuration on the fcc (111) surface has
a first-neighbour geometry  very close to that of
SbAg$_3$, the only stable ordered Sb-Ag compound  known, having
a tetragonally-distorted fcc structure\cite{vill}.

The substitutional Sb adsorbate sits in the
surface vacancy in a position very
close to the ideal fcc location of the substituted Ag atom,
with an outward relaxation of only  5 to 8 \% of the interlayer
spacing, {\it ie.} about 0.25-0.35 \AA.
Due to the effective in-plane screening thus provided by the  surrounding
substrate atoms, the substitutional Sb adatoms interact only
weakly with each other; the  adsorption energy for the substitutional site
 does not  change much  at low coverage if  the local environment for
substitutional Sb is  conserved (see Table \ref{tab1}).
If we  assume the $\Theta$=1/4 adsorption
energy to be the low coverage limit value, and the
coverage to be  low enough, the entropic contribution to
the free energy can  overcome the internal energy difference at
relatively low temperatures. At a coverage of $\Theta\simeq$0.1,
the annealing temperature (say, 600 K) is  sufficient to cause disordering
with respect to the $\sqrt{3}$$\times$$\sqrt{3}$ arrangement.
If  the substitutional adsorption is activated by annealing, we
therefore expect that
the  substitutional surface alloy thus obtained will be disordered.

To clarify the effects of Sb adsorption on the
growth mode of Ag, we studied  Ag adsorption on clean and Sb-covered Ag (111).
We used the 2$\times$2 cell for these studies, both because
neighboring adsorbates are reasonably decoupled from   each other, and
because Ag can be adsorbed  on the substitutional Sb-covered surface
 either as a nearest neighbour to Sb, or not. We call these two sites ``near''
and ``far''.
The adsorption energies are summarized in Table \ref{tab3}. The main
result is that Ag has a higher adsorption energy on clean Ag  than at
both of the sites on Sb-covered Ag. Among the latter sites, the
``near'' site is marginally disfavored, and it would be probably more
so if Sb had been allowed to relax outwards (see Table \ref{tab1}).

It is thus energetically preferrable for Ag to sit on clean portions of the
surface, while the vicinity of substitutional Sb centers is unfavourable. This
could be called long-range ``site'' blocking,
as the interaction giving rise to it is apparently long-ranged.
We put ``site'' in quotes because the adsorbate potential energy
is expected to change
 gradually as the adsorbate approaches the Sb centers, so that
the average diffusion barrier for Ag  increases already at some distance from,
and not only {\it at}, the Sb centers.
Diffusion barriers for Ag on Ag (111) are smaller than 0.1 eV: near an Sb
center, they increase significantly, namely to about 0.4-0.5 eV.
As a consequence, adsorbed Sb in the substitutional configuration
reduces the surface mobility of Ag. The presence of  substitutional Sb should
 therefore favor the growth of  small-sized
Ag islands. If the island density is high, one expects that they
  coalesce into a single layer before overgrowth on the islands
 can occur, as it is generally believed that
small islands have lower energy barriers  at descending
steps. As deposited  Ag
 covers the Ag:Sb surface alloy layer, Sb atoms find themselves
in the disfavoured sublayer configuration, and will thus tend to segregate to
the  new surface layer. The alloy surface layer is thus
reestablished, and the process can start  again.

Recent STM experiments\cite{vdv2} on this system have indeed shown
that on annealed Sb-covered surfaces,
 Sb induces  a high density of small Ag islands on the surface, and
that it  efficiently segregates  upon deposition
of  Ag. Another  observation is that upon
annealing at about 550 K,   Sb is adsorbed substitutionally and,
at very low coverages, it
forms a disordered  2D array, in agreement with our prediction.
For the unannealed surface, the on-surface fcc site is  occupied
at room temperature; this agrees with our estimate of
the activation of substitutional adsorption. At very low
coverage, Sb is observed to form islands.
Our  largest calculated adsorption energy for on-surface
adsorption is that of the 2$\times$2 superstructure;
we cannot exclude however that the adsorption energy may increase further
in the extreme low-coverage limit, which is computationally very demanding and
has not been addressed here.

In summary, we presented  {\it ab initio}
calculations of Sb and Ag adsorption on clean and Sb-covered Ag (111).
For Sb, the substitutional adsorption site is energetically
highly favoured with respect to ``normal'' on-surface
sites; in addition, subsurface positions are also strongly disfavoured.
Sb is thus effectively
confined into the surface and forms a segregating surface alloy.
This alloy should disorder, at low coverages, for typical annealing
temperatures. As to Ag, we find it to be sizably more bound on clean Ag(111)
than on substitutional Sb-covered Ag: this indicates that Sb produces a
 site blocking, or more precisely, a significant increase of the
diffusion barrier for  Ag adatoms approaching the Sb centers.
Based on these results, we offered an explanation of the
recently observed Sb-induced layer-by-layer homoepitaxial growth of Ag (111):
substitutionally-adsorbed Sb induces, by mobility reduction, a  high
density of small-sized Ag islands which  coalesce into a regular 2D layer;
as Ag covers the surface, Sb segregates to the newly formed layer, thus
reestablishing the alloy layer at the surface, and the process starts
again. Most of our results seem  to be confirmed by
recent STM experiments\cite{vdv2}.

We thank R. Stumpf for helpful discussions, and J. Vrijmoet for
communicating his results prior to publication.
This work was partly supported by the Deutsche
Forschungsgemeinschaft  within  Sonderforschungsbereich 1421.

%%%%%%%%%%%%%%%%%%%%%%%%%%%%%%%
%	 references
%%%%%%%%%%%%%%%%%%%%%%%%%%%%%%%
\narrowtext

%%%%%%%%%%%%%%%%%%%%%%%%%%%%%%%%%%%%%%%%%%%%%%%%%%%%%%%%%%
%		end references
%%%%%%%%%%%%%%%%%%%%%%%%%%%%%%%%%%%%%%%%%%%%%%%%%%%%%%%%%%
\narrowtext
%
% table 1 ads energies
%
\begin{table}
\begin{tabular}{l|ccc}
\multicolumn{1}{c}{ } &
\multicolumn{1}{c}{ $\sqrt{3}\times\sqrt{3}$}&
\multicolumn{1}{c}{ 2$\times$2} &
\multicolumn{1}{c}{ 1$\times$1} \\
\tableline
E$_{\rm ad}^{\rm sub}$   & 4.49 & 4.37  & ---   \\
{\footnotesize relaxation} &  +5\% & +8\% & --- \\
\tableline
E$_{\rm ad}^{\rm fcc}$   & 3.26 & 3.34  & 3.22   \\
{\footnotesize relaxation} &  --11\% & --5\% & +6\% \\
\tableline
E$_{\rm ad}^{\rm sublayer}$   & 3.41 & 3.45  &  2.71   \\
\end{tabular}
\caption{Adsorption energies (in eV/atom)   of Sb on Ag (111)
for the adsorption sites and coverages studied here. Vertical adsorbate
relaxations compared to ideal silver fcc position (in percentage of interlayer
spacing) are also given.}
\label{tab1}
\end{table}
%
%% table 2 formation energies
%
\begin{table}
\begin{tabular}{l|cc}
\multicolumn{1}{c}{ } &
\multicolumn{1}{c}{ $\sqrt{3}\times\sqrt{3}$}&
\multicolumn{1}{c}{ 2$\times$2} \\
\tableline
E$_{\rm f}^{\rm vac}$   & 0.69  & 0.66    \\
E$_{\rm f}^{\rm Fp}$   & 1.46 & 1.43    \\
\end{tabular}
\caption{Vacancy and Frenkel pair formation energies (eV)}
\label{tab2}
\end{table}
%
%% table 3 ag adsorption energy
%
\begin{table}
\begin{tabular}{l|ccc}
\multicolumn{1}{c}{ } &
\multicolumn{1}{c}{clean} &
\multicolumn{1}{c}{far} &
\multicolumn{1}{c}{near} \\
\tableline
$E_{\rm ads}^{\rm Ag}$   & 2.41 & 2.02  & 1.99   \\
{\footnotesize relaxation} &  --9\% & --9\% & --5\% \\
\end{tabular}
\caption{Adsorption energy (eV/atom) and relaxation compared to
the ideal Ag position (in percentage of ideal interlayer spacing)
for Ag on clean and Sb-covered Ag (111).
Clean: Ag on Ag (111); far: Ag on Sb:Ag(111), ``far'' site;
near: Ag on Sb:Ag(111), ``near'' site.}
\label{tab3}
\end{table}
%%%%%%%%%%%%%%%%%%%%%%%%%%%%%%%%%%%%%%%%%%%%%%%%%%%%%%%%%%%%%%
%%%%%%%%%%%%%%%%%%%%%%%%% end tables %%%%%%%%%%%%%%%%%%%%%%%%%
%%%%%%%%%%%%%%%%%%%%%%%%%%%%%%%%%%%%%%%%%%%%%%%%%%%%%%%%%%%%%%

\begin{references}
\bibitem{surfe}
The results may be reformulated in terms of surface energy reduction caused
by adsorption. This requires  the definition of a chemical potential for the
atoms to be adsorbed.  Here we adopt the adsorption energy terminology.
\end{references}
\end{document}